\def\year{2015}
\documentclass[letterpaper]{article}
\usepackage{aaai}
\usepackage{times}

\usepackage{graphics} 
\usepackage{txfonts}
\usepackage{times}    
\usepackage[pdftex]{hyperref}
\usepackage{color}
\usepackage{textcomp}
\usepackage{booktabs}
\usepackage{ccicons}
\usepackage{todonotes}
\usepackage{floatrow}
\usepackage{balance}
\usepackage{comment}

\usepackage{helvet}
\usepackage{courier}

\begin{document}
%
\title{Towards Generating Virtual Movement from Textual Instructions\\ {\normalsize A Case Study in Quality Assessment}}
\author{Himangshu Sarma, Robert Porzel, Jan Smeddinck, Rainer Malaka\\
Digital Media Lab, TZI, University of Bremen\\
Bibliothekstr. 1\\
28359 Bremen, Germany\\
\{sarma, porzel, smeddinck, malaka\}@tzi.de
}
\maketitle

\begin{abstract}
\begin{quote}Many application areas ranging from serious games for health to learning by demonstration in robotics, could benefit from large body movement datasets extracted from textual instructions accompanied by images. The interpretation of instructions for the automatic generation of the corresponding motions (e.g. exercises) and the validation of these movements are difficult tasks. In this article we describe a first step towards achieving automated extraction. We have recorded five different exercises in random order with the help of seven amateur performers using a Kinect. During the recording, we found that the same exercise was interpreted differently by each human performer even though they were given identical textual instructions. We performed a quality assessment study based on that data using a crowdsourcing approach and tested the inter-rater agreement for different types of visualizations, where the RGB-based visualization showed the best agreement among the annotators.
\end{quote}
\end{abstract}

\section{Introduction}

Assessing the quality of human body movement performances is an important task in many application areas, ranging from sports to therapy, learning by demonstration in robotics, automated systems for generative animation, and many more. For example, the manual transformation of physical therapy exercises into computer-supported playful exercises in the form of so-called exergames or levels of exergames requires a lot of time and effort, making it impractical for therapists or smaller practices to transform their preferred sets of therapeutic exercises into exergames to be used by their patients. Motivated by our use-case of automatically generating movement patterns to be used in motion-based games for the support of physiotherapy, rehabilitation, and prevention, we thus set out to explore the potential of crowd-based quality of motion assessments, as a necessary intermediate step in the extraction and validation of motions. The human-computation approach is promising in this regard, since the task involves many aspects that are easy for humans, but difficult for machines~\cite{krause2011human}. Since it is known that even human experts in quality of movement judgements share little inter-rater agreement~\cite{pomeroy2003agreement}, we set out to explore whether it is possible to achieve a level of inter-rater reliability that would even allow for quality of motion-assessment, if a later cross-validation is projected and to explore which type of a motion-visualization would support the best inter-rater reliability, whereby we hypothesized that the video-based modality would yield the highest inter-annotator reliability. With this work we contribute to human computation by exploring the novel area of quality of motion assessment, where successful human computation could prove beneficial to a large number of application scenarios, and we address the relevant related independent variable of motion visualization.

\section{State of the Art}
Motion-based games for health are subject to a growing body of research and development. In a series of studies, ~\cite{uzor} have shown that these playful tools can provide a number of benefits compared to traditional instruction by exercise sheets, especially when used to augment unsupervised exercising at home. We summarize these areas to be motivation (to perform repetitive exercises), feedback (regarding the current exercise execution and summarizing developments), and customization (by manual adaptations of automatic adaptivity)~\cite{smeddinck2015exergames}. Such games can be created in a modular fashion, where the specific exercises to be supported are arbitrary, yet require manual effort for a successful implementation. There are thousands of different exercises employed by different practices, thus automated extraction methods could provide a great benefit to this area.
Furthermore, the reliable objective assessment of quality of motion, even when supervised by a therapist is a challenge, since inter-rater variance is notably high~\cite{pomeroy2003agreement}. Again, automated, or human computation supported methods could be of great benefit in this area.
Both the generation of movements from textual descriptions that are accompanied by images, as well as the validation of movement quality lend themselves as tasks for human computation, since they involve most of the typical strongholds of human computation, including intuitive decisions, aesthetic judgement, contextual reasoning, and embodiment issues~\cite{krause2011human}.

Based on the current state of the art, we set out to establish a human computation based pipeline for extracting validated movements from instruction sheets, with the goal to then explore the potential of further automating the different steps involved in that pipeline, starting with a focus on the step of quality of motion assessment.

\section{Data Collection and Results}

At the early stage of the work we developed a Physical Exercise Instruction Sheet Corpus (PEISC) of around 1000 physical exercise instructions drawn from a number of publicly available databases. On the basis of different bodily actions we categorized it into different categories such as standing, seating. For our first case study, we chose five exercises (Table~\ref{bp}) which do not require any additional equipment.

Using a Kinect device, we recorded five exercises from seven participants (3 male and 4 female; 15 to 35 years of age, M=25, SD=5). Ten iterations of every exercise were recorded from each participant in a random order. During the recording of those exercises we only provided instruction sheets and asked the participants to perform their interpretation of the exercises without any priming regarding how to perform them. We also collected basic demographic data and after every exercise we collected responses regarding the comprehensibility of the instruction sheets. Analyzing the answers from the questionnaires, we found that the exercise instruction sheets were difficult to understand for some yet seemed easy for others. Furthermore, the same exercise was sometimes performed differently by the participants. With these findings we can claim that instruction sheets are not the optimal way to instruct people to perform exercises.

We have developed 4 different categories of videos from the collected Kinect data (i.e., RGB, Depth, Skeleton, Virtual Reality) and developed a survey application, aiming to crowdsource the assessment of the quality of exercise executions and to determine the best visualization modality for high inter-rater agreement. Following the quality assessment survey, we provided a questionnaire to gather comparative responses regarding the best visualization type, movement quality of different body parts during the performance of the exercises, and to acquire additional demographic data.

In the survey, we asked the participants to read each instruction sheet followed by asking them to watch the videos of all 7 performances in all 4 categories, where all exercises and categories appeared in a random order on the screen. The participants' task was to delete the worst one and repeat that procedure until the best one remained. In total, 20 participants took part in the survey and questionnaire.

\subsection{~~Results} 
\begin{figure}[h]
										 \begin{floatrow}
											 \ffigbox{\includegraphics[width=0.9\columnwidth]{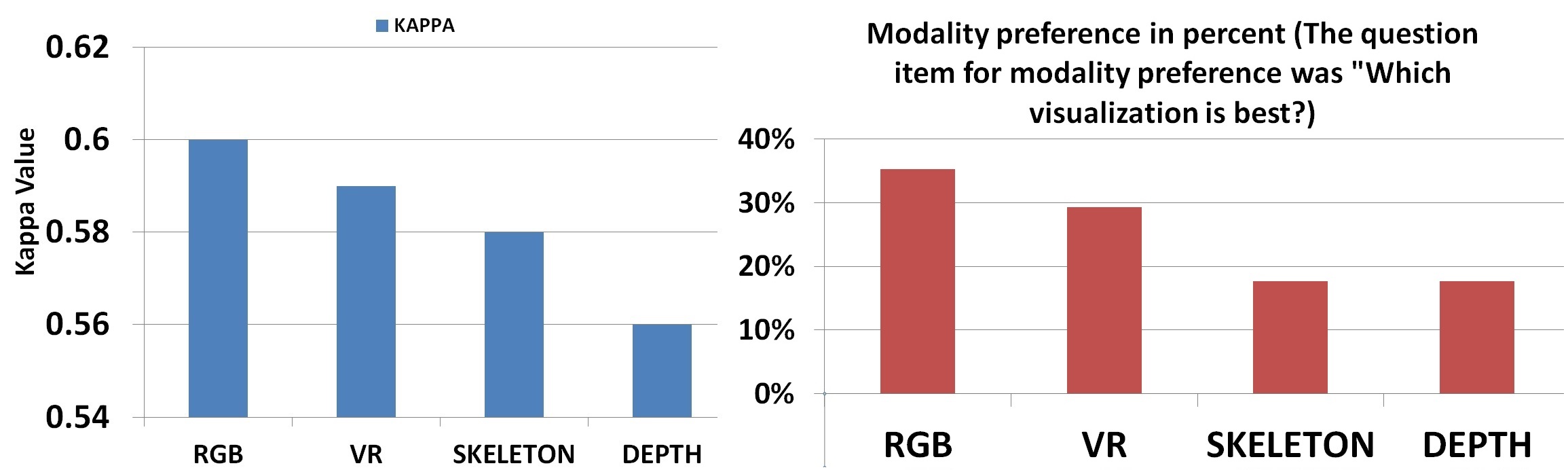}}{\tiny \caption{\tiny Agreement for different visualizations}\label{fig:kappa}}
										 \end{floatrow}
        \end{figure}

With the help of Kappa statistics~\cite{Carletta}, we calculated the best performer of all 5 exercises (shown in Table~\ref{bp}) and the best visualization type (displayed in Figure~\ref{fig:kappa}). 

\begin{table}[h]
\centering
\caption{\tiny Best performer per exercise and the inter-rater agreement on the positioning}
\label{bp}
\begin{tabular}{|l|l|l|l|l|l|}
\hline
\multicolumn{1}{|l|}{{\bf {\small Exercise}}} & {{\bf {\small Performer}}} & {{\bf {\small Kappa}}} \\
\hline
{\footnotesize Squats}      & {\footnotesize Performer 1}         & {\footnotesize 0.51}                 \\
{\footnotesize Lateral Lungs}      & {\footnotesize Performer 2}        & {\footnotesize 0.59}                 \\
{\footnotesize Standing IT}      & {\footnotesize Performer 6}        & {\footnotesize 0.53}                 \\
{\footnotesize Forward Lungs}      & {\footnotesize Performer 1}         & {\footnotesize 0.74}                 \\
{\footnotesize Reverse Lunges}      & {\footnotesize Performer 6}         & {\footnotesize 0.57}                 \\
\hline
\end{tabular}
\end{table}

\section{Conclusion and Future Work}
In this paper, we have presented a precursory step towards an automated pipeline to go from texts to virtual motions. Here, we have shown a case study for five different exercises performed by seven different humans with the help of typical exercise instruction sheets. Using a crowdsourcing approach, we have found that assessing the quality of the performed exercises is not an easy task for humans and that the RGB-type (regular video) of visualization yields the most reliable ratings, which could be expected, since it was the visualization modality that subjects were likely most familiar with. In the future, we will aim at developing an automated system that produces virtual motions based on typical exercise instruction sheets as an input. A follow-up to the study presented in this abstract will help inform the decision on the visualization modality that will be employed for those automatically generated exercise executions for an intermediate judgement of the quality of these generated executions. We are hoping to use insights from that step to, in turn, inform the further automation of the overall pipeline.

%
%
%
%
%
\balance{}

\bibliographystyle{aaai}
\bibliography{sample}

\end{document}